\documentclass
[
    ,final            
  ]
  {aipproc}
\usepackage{epsfig}
\usepackage{graphics}
\layoutstyle{8x11single}

\begin{document}

\title{Nucleation of antikaon condensed matter in proto neutron stars}

\classification{26.60.-c, 97.60.Jd, 05.70.Ce,  82.60.Nh}
                
\keywords      {Neutron stars, antikaon condensation, nucleation}

\author{Sarmistha Banik}
{
  address={Saha Institute of Nuclear Physics, 1$/$AF Bidhan Nagar, Kolkata -7000064, India.}
}

\author{Rana Nandi}
{
  address={Saha Institute of Nuclear Physics, 1$/$AF Bidhan Nagar, Kolkata -7000064, India.}
}

\begin{abstract}
A first order phase transition from nuclear matter to antikaon condensed matter
may proceed through thermal nucleation of a critical droplet of antikaon 
condensed matter during the early evolution of proto neutron stars (PNS). 
Droplets 
of new phase having radii larger than a critical radius would survive and 
grow, if the latent heat is transported from the droplet surface to the 
metastable phase.  We investigate the effect of shear viscosity on the thermal 
nucleation time of the droplets of antikaon condensed matter. In this connection
we particularly study the contribution of neutrinos in the shear viscosity 
and nucleation in PNS.
\end{abstract}

\maketitle


\section{Introduction}
A first order phase transition from hadronic to exotic matter phases may
proceed through the nucleation of droplets of the new phase. The formation
of droplets of quark matter \citep{Bombaci1, Mintz, Bombaci2} and antikaon 
condensed matter  \citep{Norsen, NorsenReddy} in neutrino-free neutron stars (NS) was 
studied using the homogeneous nucleation theory of Langer \citep{Langer}. 
Droplets of new phase may appear in the metastable nuclear matter due to thermal 
fluctuations. The droplets of the stable phase with radii larger than a 
critical radius survives and grows if the latent  heat is transported 
from the surface of the droplet to the metastable state. This heat 
transportation occurs through thermal dissipation \cite {Las}
and viscous damping \citep{Raj}.

We have seen that the onset of antikaon condensate influences the shear 
viscosity of NS matter composed of neutron(n), proton(p), electron(e) and 
muon($\mu$) that can interact by strong or electromagnetic interactions 
\citep{shear}.  Effect of shear viscosity on the nucleation
of the antikaon condensed matter was recently been studied \citep{nucleat}
in deleptonised NS, after the neutrinos are emitted. 
Here we investigate the contribution of neutrinos to the shear viscosity 
and nucleation of antikaon condensates.
In the PNS where the temperature is of the order of a few 10s 
of MeV, neutrinos are trapped because their mean free paths under these 
conditions are small compared to the radius of the star. On the other hand, 
they are very effective at transporting both heat and momentum because their 
mean free paths are orders of magnitude larger than that for other particles.

\section{Formalism}
We adopt the homogeneous nucleation theory of Langer \citep{Langer} to calculate
the thermal fluctuation rate for a first order phase transition from the
charge-neutral and beta-equilibrated nuclear matter to $K^-$ condensed matter
in a neutrino-trapped PNS. The thermal nucleation rate is 
given by 
\begin{equation}
 I= \Gamma_0 \exp(-\frac{\Delta F (R_c)}{T}),
\end{equation}
where $\Delta F= $ is the change in free energy required to activate the
formation of  the critical droplet. $\Gamma_0=\frac{\kappa}{2 \pi}\Omega_0$, is the prefactor of which  $\Omega_0=\frac{2}{3\sqrt{3}} \left(\frac {\sigma}{T}\right)^{3/2}
\left(\frac {R_C}{\xi}\right)^4~$ is the statistical prefactor and
$\kappa=\frac{2 \sigma}{R_c^3 (\Delta w)^2}\left[\lambda T+2 \left(\frac{4}{3} \eta+\zeta \right)\right]$ 
is the dynamical prefactor. Here $\sigma$ is the surface tension for the 
surface separating the two phases, $\xi$ is the correlation length for kaons, 
$\Delta w$ is the difference of the enthalpy of the two phases, $\lambda$ the 
thermal conductivity, $\eta$ and $\zeta$ are the shear and bulk viscosity 
respectively. The free energy is maximum at this critical radius given by
\begin{equation}
R_c(T)=\frac{2 \sigma}{\left(P^K-P^N\right)}.
\end{equation}
Finally the thermal nucleation time is given by $\tau_{th}=(V_{nuc}I)^{-1}$,
where $V_{nuc}=\frac {4 \pi}{3}R_{nuc}^3$ is the volume of the  
core, where the thermodynamic variables is assumed to remain constant.  

To calculate shear viscosity in the PNS, which is
mostly contributed by the neutrinos \citep{Pethick}, we consider the scattering 
of neutrinos $\nu_e + N \rightarrow \nu_e + N $ where N=n, p, e. 
Shear viscosity of neutrinos due to scattering is calculated using the 
coupled Boltzmann transport equation \citep{Pethick}. For the deleptonised 
NS, total shear viscosity is given by $\eta=\eta_n+\eta_p+\eta_e+\eta_{\mu}$ as 
in Ref. \citep{shear}.
%
\paragraph{The Model EoS}

In order to calculate the shear viscosity and critical radius ($R_c$), we need 
to know the  EoS,
that we construct at finite temperature using the relativistic mean field 
model \citep{Pons, crit}. The interaction 
between baryons is mediated by the exchange of scalar ($\sigma$) and vector 
($\omega,\rho$) mesons. This picture is consistently extended to include the 
kaons. We use the parameter sets of GM1 model \citep{GM} for nucleon-meson 
coupling constants. The kaon-meson coupling constants are determined using
quark model, isospin counting rule and the real part of $K^-$ potential depth, 
that we take as -160 MeV in our calculation \citep{shear}.

\subsection{Results}

\begin{figure}[t]
\includegraphics[height=0.3\textheight, width=0.79\textwidth]{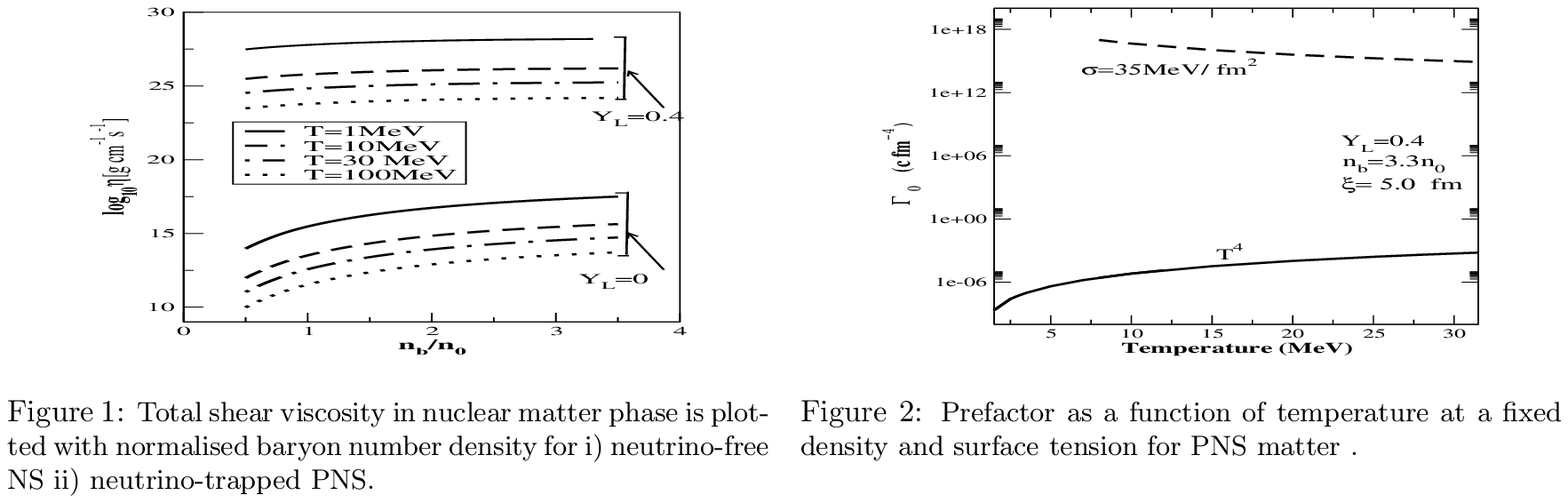}
\end{figure}
The total shear viscosity is shown as a 
function of normalised baryon density for different temperatures in Fig. 1. 
We consider two cases i)the deleptonised NS matter, where the total shear 
viscosity has contribution from all the species such as n, p, e and $\mu$;
ii)for the neutrino-trapped PNS matter (lepton fraction 
$Y_L=0.4$), where the major contribution comes from neutrinos. In 
both the cases shear viscosity decreases with rising temperature.

The prefactor $\Gamma_0$ is plotted as a function of temperature for PNS 
in Fig. 2 and is compared when it is approximated by $T^4$ from dimensional 
analysis. 
Here we find the shear viscosity term changes the prefactor by a large order of 
magnitude compared to $T^4$ approximation. This difference is much more 
pronounced in PNS compared to NS \citep{nucleat}.

In Fig. 3 we display the nucleation time as a function of temperature for 
a set of values of surface tension at a fixed baryon density for 
lepton-trapped PNS matter and find that both the droplet radii and the thermal 
nucleation time strongly depend on the surface tension.   
Nucleation may not occur in PNS for surface tension <30 MeV fm$^{-2}$ as
the radius of droplet is less than the correlation length ($\xi=\sim$ 5 fm). 
We approximate that the radius of the droplet should be greater than $\xi$
in this calculation\citep{NorsenReddy,Las}. For NS case, nucleation is observed
to be possible for $\sigma < 20$ MeV fm$^{-2}$ \citep{nucleat}. Larger 
viscosity there leads to larger value of T which might melt the condensate.

\begin{figure}
\includegraphics[height=0.3\textheight, width=0.79\textwidth]{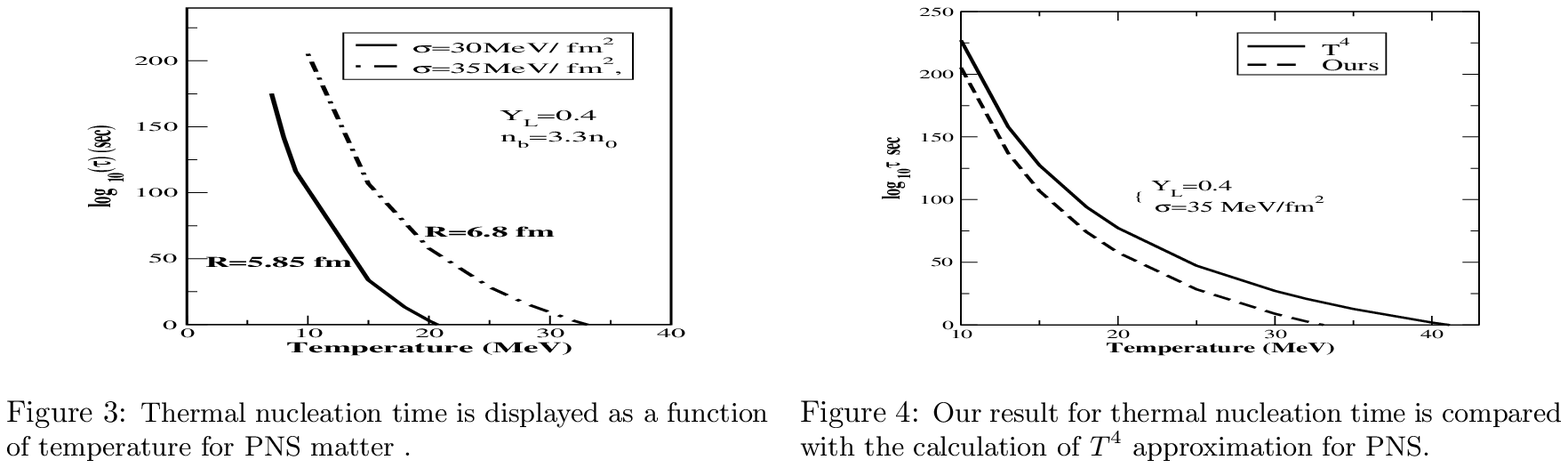}
\end{figure}
Finally in Fig. 4 we compare the results of thermal nucleation time taking 
into account the effect of shear viscosity in the prefactor with that of the 
prefactor approximated by $T^4$. 
For PNS we displayed the results for $n_b = 3.30n_0$ and $\sigma =35$MeV fm$^{-2}$. We find the result of the $T^4$ approximation a few orders 
of magnitude higher than those of our calculation. These results demonstrate 
the importance of including the shear viscosity in the prefactor of Eq. (1) 
in  the calculation of thermal nucleation time. We already obtained similar 
results for NS matter \citep{nucleat}. Also, it may be mentioned that 
nucleation of antikaon condensates is possible if $\tau_{th} < \tau_{cooling}
(\sim 100s)$. That is possible for PNS with $\sigma \geq 30$ MeV fm$^{-2}$.

\subsubsection{Summary}
We have investigated the role of shear viscosity  on the thermal nucleation rate
for the formation of a critical droplet of antikaon condensed matter. For this we
considered a first order phase transition from the nuclear to antikaon condensed matter
in PNS. We have seen that the droplet radii increase with increasing surface tension.
We also compare the nucleation times for NS and PNS and found that nucleation is possible
for  a lower value of surface
tension $\sigma <$ 20 MeV fm$^{-2}$ in NS while it may be possible only for higher value
in PNS ( $\sigma \geq $ 30 MeV fm$^{-2}$).

\begin{theacknowledgments}
S.B. would like to acknowledge the Department of Science \& Technology, India
for the travel grant to present this paper in the conference PANIC11.
\end{theacknowledgments}


\bibliographystyle{aipproc}   

\end{document}